\documentclass[twoside]{dis07}
\usepackage[latin1]{inputenc}
\usepackage[dvips]{graphicx,epsfig,color}
\usepackage{wrapfig,rotating}
\usepackage{amssymb,amsmath,array}

\newcommand{\pmin}{p_{\mbox{\footnotesize{min}}}}

\def\lsim{\,\lower.25ex\hbox{$\scriptstyle\sim$}\kern-1.30ex%
\raise 0.55ex\hbox{$\scriptstyle <$}\,}

\newcommand{\ej}{\mbox{$e$-$j$}}

\newcommand{\mujnp}{\mbox{$\mu$-$j$-$\nu$}}

\newcommand{\jjjj}{\mbox{$j$-$j$-$j$-$j$}}

\newcommand{\jjjjj}{\mbox{$j$-$j$-$j$-$j$-$j$}}
\newcommand{\jjjjnp}{\mbox{$j$-$j$-$j$-$j$-$\nu$}}

\begin{document}
\title{A General Search for New Phenomena at HERA}

\author{E. Sauvan\\
\uppercase{O}n behalf of the \uppercase{H}1 \uppercase{C}ollaboration
%
%
\vspace{.3cm}\\
%
CPPM, IN2P3-CNRS et Universit\'e de la M\'editerran\'ee\\ 
163 Avenue de Luminy F-13288 Marseille, France\\
}

\maketitle

\begin{abstract}
  A model-independent search for
  deviations from the Standard Model prediction is performed
  in $e^+ p$ and $e^- p$ collisions at HERA~II using all high energy data recorded by the H1 experiment. This 
  corresponds to a total integrated luminosity of $337$~pb$^{-1}$.
  All event topologies 
  involving isolated electrons, photons, muons, neutrinos and jets with
  high transverse momenta are investigated in a single analysis.
  Events are assigned to exclusive classes according to their
  final state.
  A statistical algorithm is used to search for
  deviations from the Standard Model in distributions of the scalar sum of
  transverse momenta or invariant mass of final state particles and to quantify their significance.
  A good agreement with the Standard Model prediction is observed in most
  of the event classes.
  The most siginificant deviation is found in the \mujnp~channel in $e^+p$ collisions.
\end{abstract}

\section{Introduction}
At HERA electrons\footnote{
  In this paper ``electrons'' refers to both electrons and positrons, unless otherwise stated.}
and protons collide at a centre-of-mass energy of up to $319$~GeV. 
These high-energy electron-proton interactions provide a 
testing ground for the Standard Model (SM) complementary to $e^+e^-$ and $p\overline{p}$ scattering. 
The approach presented here consists of a comprehensive
and generic search for deviations from the SM prediction at large 
transverse momenta. The present analysis follows closely the strategy of the previous publication from the H1 experiment~\cite{Aktas:2004pz}.
All high $P_T$ final state configurations involving
electrons ($e$), muons ($\mu$), jets ($j$), photons ($\gamma$) or neutrinos ($\nu$) are systematically 
investigated. 
The complete HERA II data sample ($2003$--$2007$) is used, corresponding 
to a total integrated luminosity of $337$~pb$^{-1}$ shared between $e^+p$ ($178$ $\mbox{pb}^{-1}$) and $e^-p$ ($159$ $\mbox{pb}^{-1}$) collisions.


\section{Data Analysis and Results}

All final states containing at least two objects ($e$, $\mu$, $j$, $\gamma$, $\nu$) with 
$P_T >$~$20$~GeV in the polar angle
range  $10^\circ < \theta < 140^\circ$ are investigated. 
All selected events are classified into exclusive event classes 
according to the number and types of objects detected in the final state 
(e.g.  \ej, \mujnp, \jjjjj). 
The criteria used in the identification of each type of particle are chosen to ensure an unambiguous identification, while retaining high efficiencies~\cite{Aktas:2004pz}.
All experimentally accessible combinations of objects have been studied and data events are found in $23$ event classes.

A precise and reliable estimate of all relevant processes present at high transverse momentum in $ep$ interactions is needed to ensure an unbiased comparison to the SM.
Hence several Monte Carlo generators are used to generate a large number of events in all event classes, carefully avoiding double-counting of processes. The simulation contains the order $\alpha_S$ matrix elements for QCD processes, while second order $\alpha$ matrix elements are used to calculate QED processes. Additional jets are modelled using leading logarithmic parton showers as a representation of higher order QCD radiation.

\begin{figure}[htbp]
\begin{center}
\includegraphics[width=0.55\columnwidth,angle=-90]{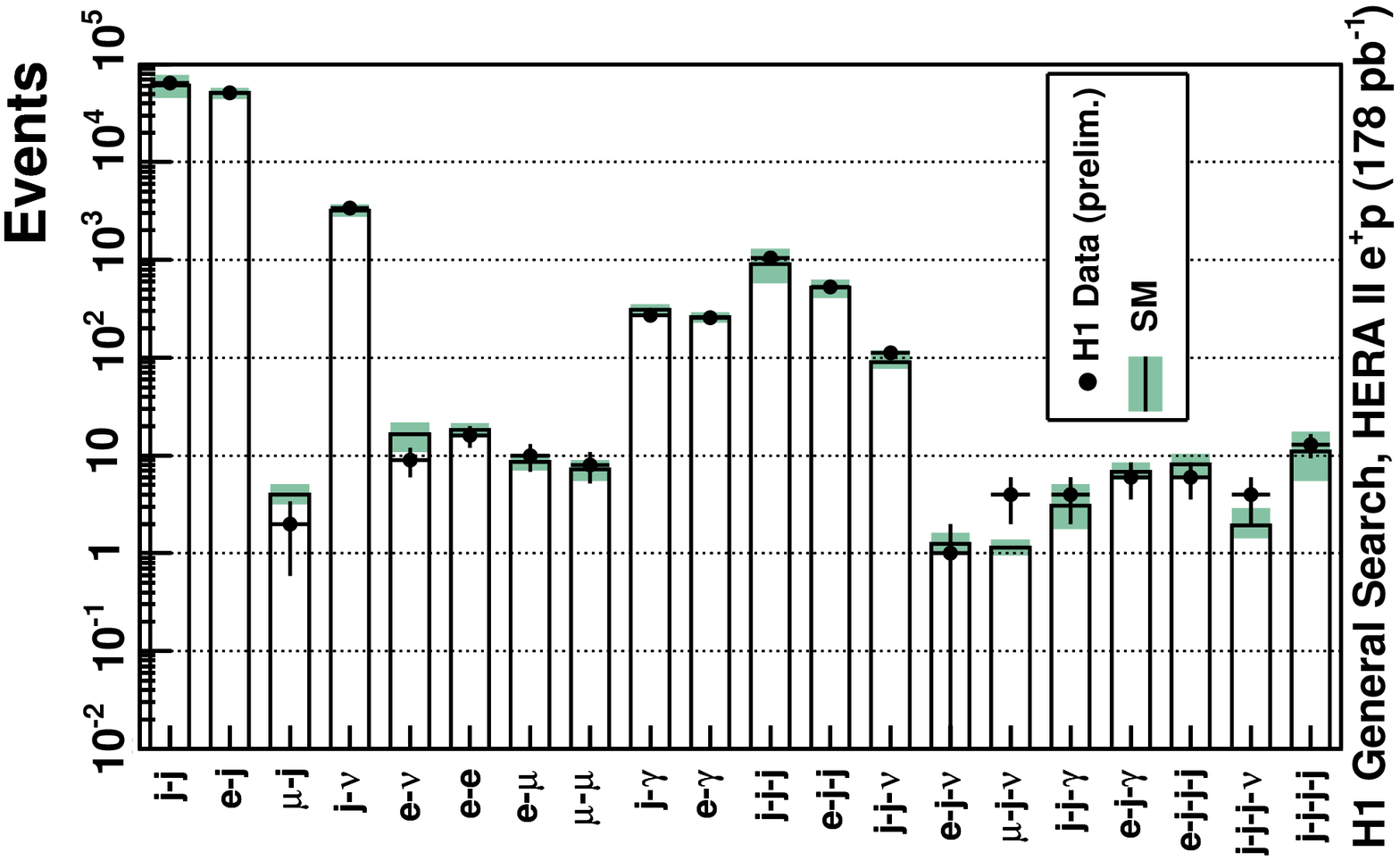}\put(-30,-160){{\bf (a)}}
\includegraphics[width=0.55\columnwidth,angle=-90]{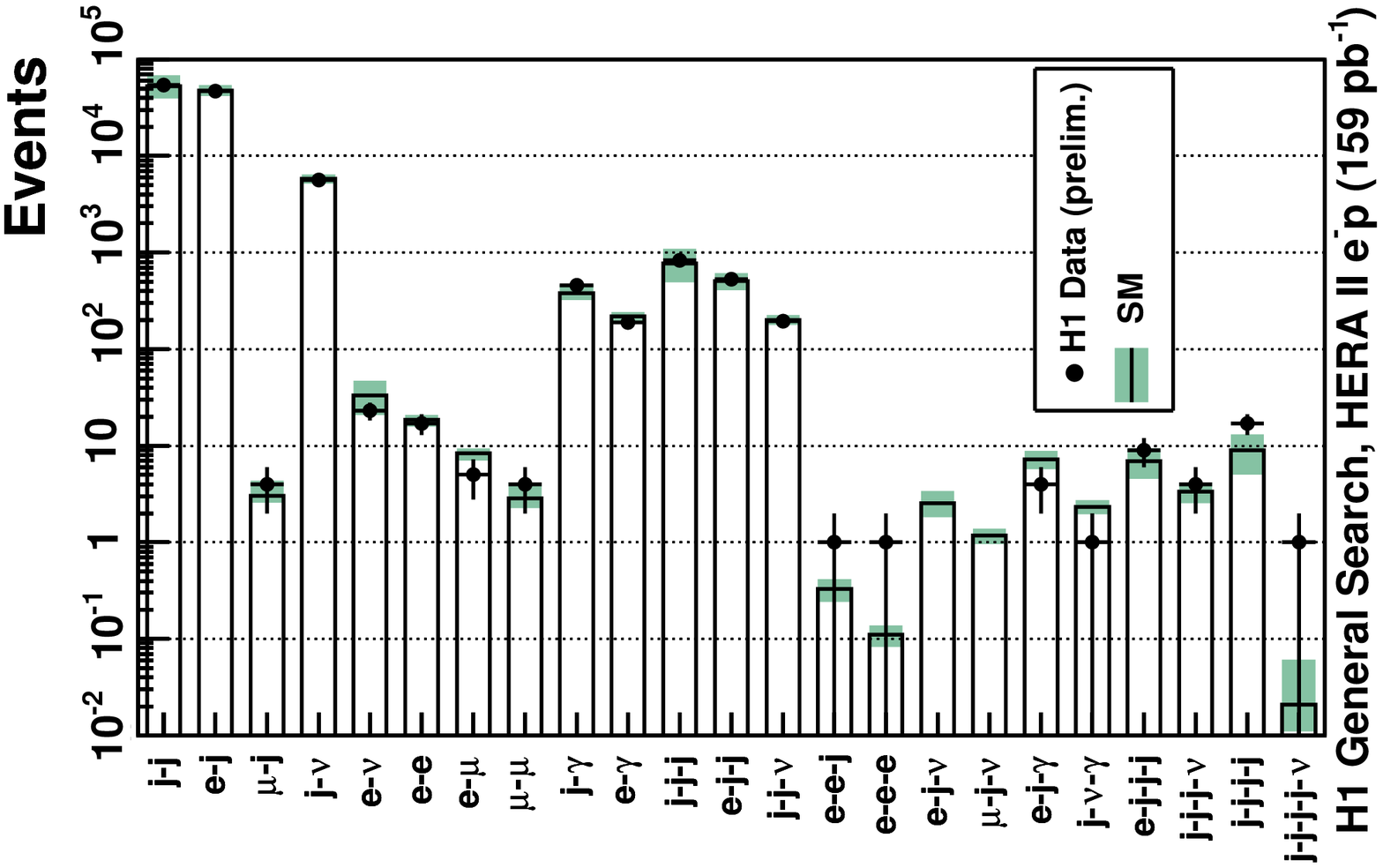}\put(-30,-160){{\bf (b)}}
\caption{The data and the SM expectation for all event classes 
    with observed data events or a SM expectation greater than one event. The results are presented separately for $e^+p$ (a) and $e^-p$ (b) collisions.
    }\label{fig:yields}
\end{center}
\end{figure}

\begin{figure}[htbp]
\begin{center}
\includegraphics[width=0.7\columnwidth]{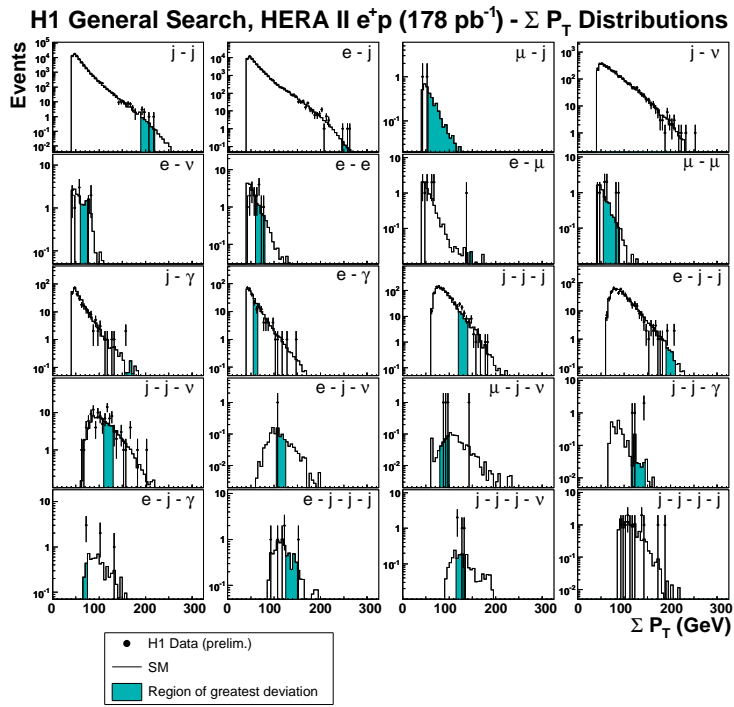}
\vspace*{-20pt}
\caption{Distributions of $\sum P_T$ for classes with at least one data event, for $e^+p$ data. 
    The shaded areas show the regions of greatest deviation 
    chosen by the search algorithm.}\label{fig:SPt}
\end{center}
\end{figure}

The event yields observed in each event class are presented and compared to the SM expectation
in figures~\ref{fig:yields}(a) and (b) for $e^+p$ and $e^-p$ collisions, respectively. 
%
In each class, a good description of the number of observed data events by the SM prediction is seen.
This demonstrates the good understanding of the detector response and of the SM processes in the considered phase space.
Distributions of the scalar sum of transverse momenta $\sum P_T$ of all objects are presented in figure~\ref{fig:SPt} for $e^+p$ data.

\section{Search for deviations}
In order to quantify the level of agreement between 
the data and the SM expectation and to identify regions of possible 
deviations, the same search algorithm as developed in~\cite{Aktas:2004pz} is used.
All possible regions in the histograms of $\sum P_T$ and $M_{all}$ distributions are considered. 
The number of data events ($N_{obs}$), the SM 
expectation ($N_{SM}$) and its total systematic uncertainty ($\delta N_{SM}$) are calculated for each region.
A statistical estimator $p$ is defined to judge which region is of 
most interest. This estimator is derived from the convolution of the
Poisson probability density function (pdf) to account for statistical 
errors with a Gaussian pdf, $G(b;N_{SM},\delta N_{SM})$, with mean $N_{SM}$ and width $\delta N_{SM}$, to include the effect of 
non negligible systematic uncertainties~\cite{Aktas:2004pz}. 
The value of $p$ gives an estimate of the probability of a fluctuation of the SM expectation upwards (downwards) to at least (at most) the observed number of data events in the region considered.
The region of greatest interest (of greatest deviation) is the region having the smallest $p$-value, $\pmin$.

The possibility that a fluctuation with a value $\pmin$ occurs
anywhere in the distribution is estimated. 
%
%
This is achieved by creating hypothetical data histograms following the pdfs of the SM expectation. 
The algorithm is then run on those hypothetical histograms to find the region of 
greatest deviation and the corresponding $\pmin^{SM}$ is calculated.
The probability $\hat{P}$ is then defined as the 
fraction of hypothetical data histograms with a $\pmin^{SM}$ equal to or smaller than the $\pmin$ value obtained from the real data.
$\hat{P}$ is a measure of the statistical significance of the deviation observed in the data.
The event class of 
most interest for a search is thus the one with the smallest $\hat{P}$ value.
Depending on the final state, a $\pmin$-value of $5.7 \cdot 10^{-7}$ (``$5\sigma$'') corresponds 
to a value of $- \log_{10}{\hat{P}}$, the negative decade logarithm of $\hat{P}$, between $5$ and $6$. 
The overall degree of agreement with the SM can further be quantified by 
taking into account the large number of event classes studied in this analysis.
Among all studied classes there is some chance that small $\hat{P}$ values occur. 
This probability can be calculated with MC experiments. 
A MC experiment is defined as a set of hypothetical data histograms following the SM expectation with an integrated luminosity equal to the amount of data recorded.
The complete search algorithm and statistical analysis are
applied to the MC experiments analgously as to the data. 
This procedure is repeated many times.
The expectation for the $\hat{P}$ values observed in the data is then given by the 
distribution of $\hat{P}^{SM}$ values obtained from all MC experiments.
The probability to find in the MC experiments a $\hat{P}$ value smaller than in the data can be calculated and gives us the global significance of the observed deviation.


\begin{figure}[htbp]
\begin{center}
\includegraphics[width=0.42\columnwidth]{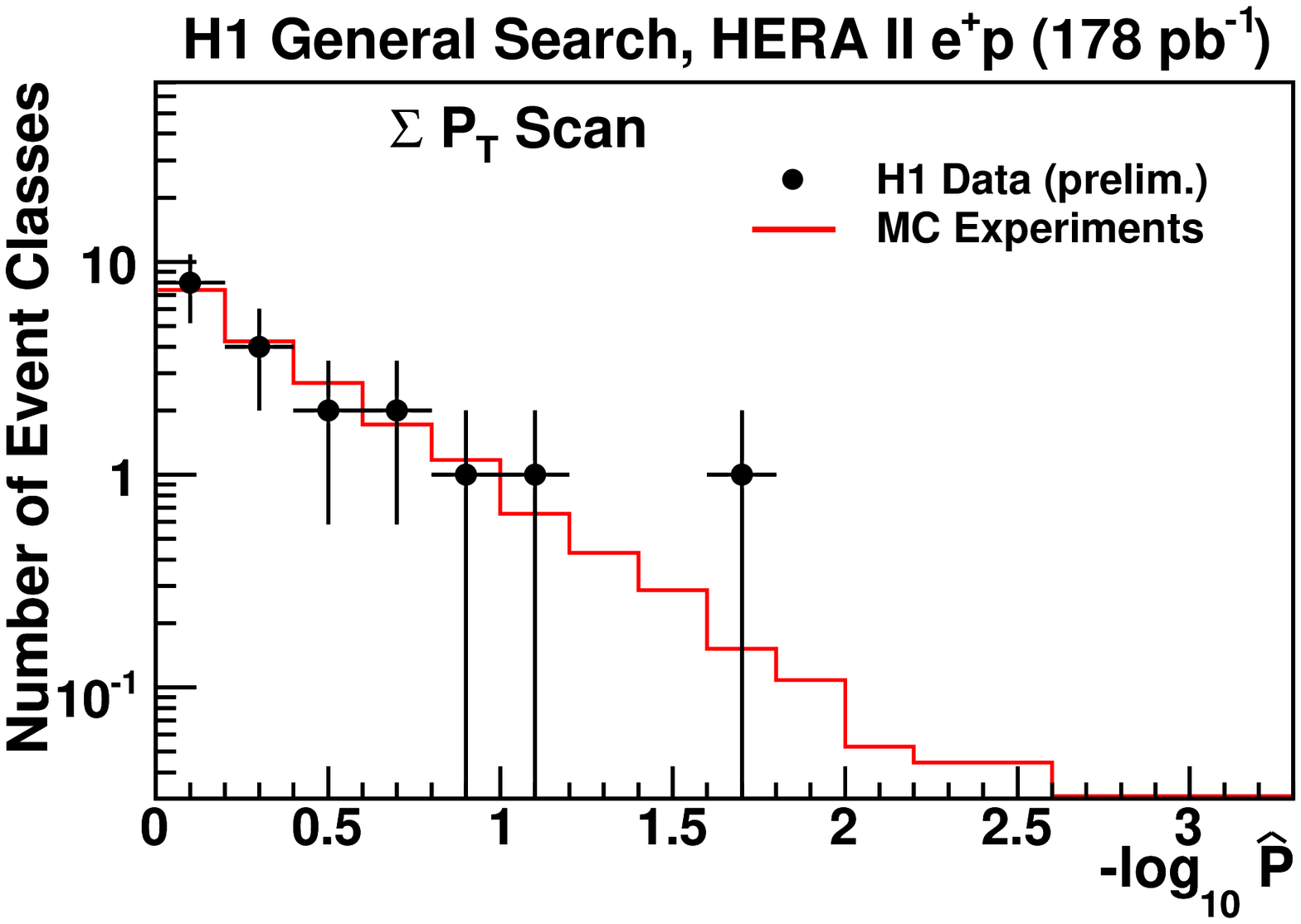}\put(-30,70){{\bf (a)}}
\includegraphics[width=0.42\columnwidth]{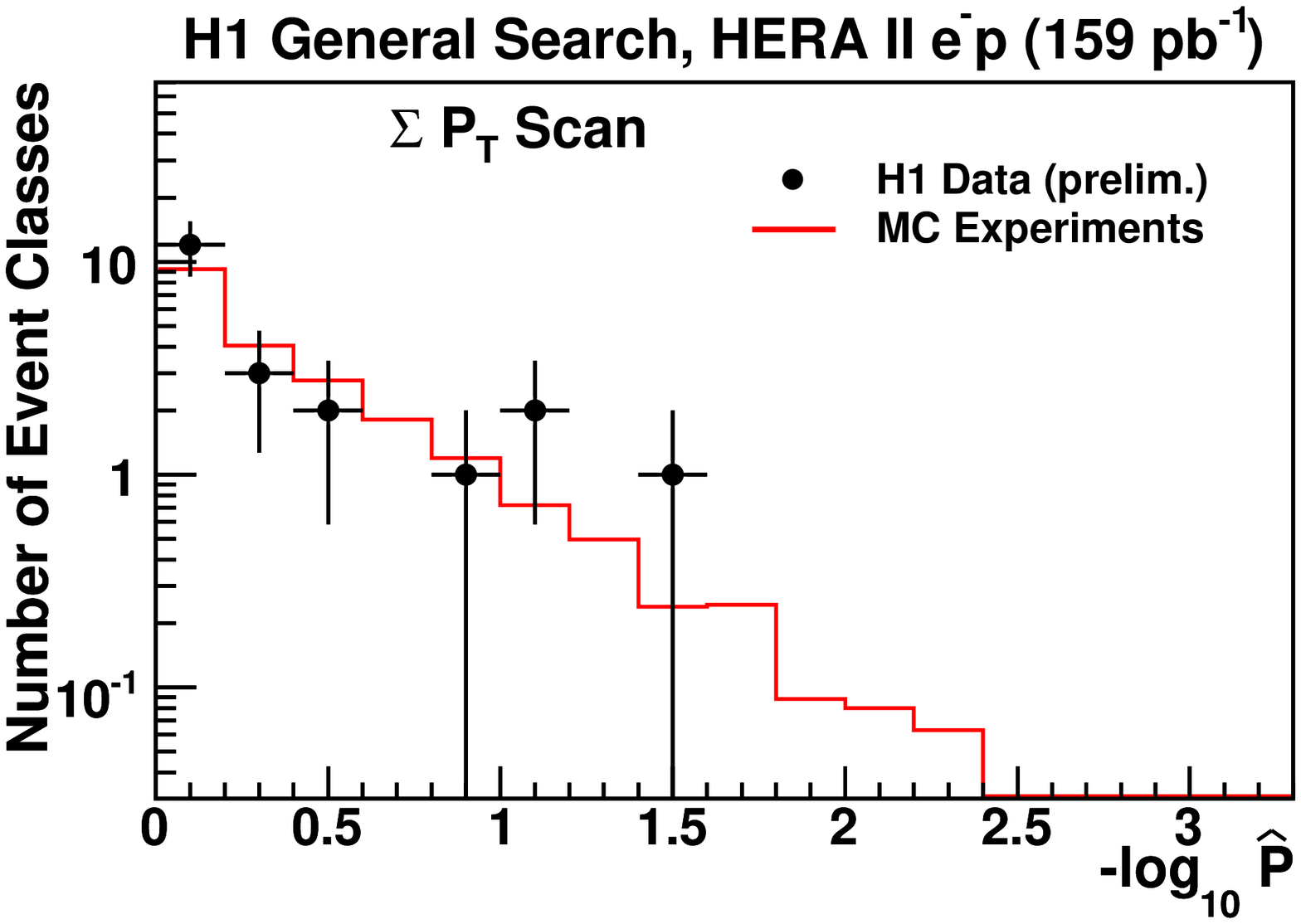}\put(-30,70){{\bf (b)}}
\vspace*{-10pt}
\caption{The $-\log_{10}{\hat{P}}$ values for the data event classes and the 
    expected distribution from MC experiments as derived by investigating
    the $\sum P_T$ distributions in $e^+p$ (a) and $e^-p$ (b) data.}\label{fig:phat}
\end{center}
\end{figure}

The $\hat{P}$ values observed in the real data in all event classes\footnote{Due to the uncertainties of the SM prediction in the \jjjj~and \jjjjnp~event classes at highest $M_{all}$ and $\sum P_T$ (see~\cite{Aktas:2004pz}), where data events are observed, no reliable $\hat{P}$ values can be calculated for these classes. These event classes are not considered to search for deviations from the SM in this extreme kinematic domain.} 
 are compared in figure~\ref{fig:phat}  to
the distribution of $\hat{P}^{SM}$ obtained from the large set of MC experiments, normalised to one experiment. The comparison is presented for the scans of the $\sum P_T$ distributions.
All $\hat{P}$ values range from $0.01$ to $0.99$, corresponding 
to event classes where no significant discrepancy between data and the SM expectation is observed. 
These results are in agreement with the expectation from MC experiments.
The most significant deviation from SM predictions is observed in the \mujnp~event class and in $e^+p$ collisions with a value of $-\log_{10}{\hat{P}}$ equal to $1.7$. In the previous H1 analysis \cite{Aktas:2004pz} based on HERA~I data and dominated by $e^+p$ collisions, the largest deviation was also found in this event class, with  $-\log_{10}{\hat{P}}=3$.

\section{Conclusions}
All the data collected with the H1 experiment during HERA~II running period ($2003$--$2007$) 
have been investigated for deviations from the SM prediction at high transverse 
momentum. All event topologies involving isolated electrons, photons, 
muons, neutrinos and jets are investigated in a single analysis.
A good agreement between data and SM expectation is found 
in most event classes.
In each event, class the invariant mass and sum of transverse momenta 
distributions of particles have been 
systematically searched for 
deviations using a statistical algorithm. 
No significant deviation is observed in the phase-space and in the event topologies covered by this analysis.
The largest deviation from SM expectation is observed in the \mujnp~event class in $e^+p$ collisions.


\begin{footnotesize}




\begin{thebibliography}{99}
\bibitem{url} Slides:
\verb$http://indico.cern.ch/contributionDisplay.py?contribId=127&sessionId=9&confId=9499$

\bibitem{Aktas:2004pz}
  A.~Aktas {\it et al.}  [H1 Collaboration],
  Phys.\ Lett.\ B {\bf 602} (2004) 14
  [hep-ex/0408044].



\end{thebibliography}
%

\end{footnotesize}


\end{document}